\documentclass[11pt]{article}
\usepackage{graphicx,amsmath,amsfonts}
\def \beq  {\begin{equation}}
\def \eeq  {\end{equation}}
\def \beqar {\begin{eqnarray}}
\def \eeqar {\end{eqnarray}}
\def\sqr#1#2{{\vcenter{\vbox{\hrule height.#2pt
\hbox{\vrule width.#2pt height#1pt \kern#1pt
\vrule width.#2pt}\hrule height.#2pt}}}}

\def\la {{\langle}}
\def\ra {{\rangle}}

\def\Tr {{\rm Tr}}
\def \tr {{\rm tr}}

\def\bp {\bar p}

\def\bA {\bar{A}}

\def\bw {\bar{w}}

\def\del {\partial}
\def\bdel{\bar{\partial}}

\def\D {{\cal D}}
\def\bz {{\bar{z}}}

\def\D {{\cal D}}

\def\half{\textstyle{1\over 2}}

\newcommand{\symmat}[4]{\left[ \begin{array}{cc} #1 & #2 \\ #3 & #4 \end{array} \right]}

\newcommand{\rpl}{\mathcal{R}_+}
\newcommand{\rmin }{\mathcal{R}_-}

\newcommand{\delsp }{\delta ^{(2)}_{S^2}}

\newcommand{\mD}{\mathcal{D}}
\newcommand{\mg}{\mathcal{G}}
\begin{document}
\fontfamily{cmr}\fontsize{11pt}{15.0pt}\selectfont
\def \CMP {{ Commun. Math. Phys.}}
\def \PRL {{ Phys. Rev. Lett.}}
\def \PL {{Phys. Lett.}}
\def \NPBProc {{ Nucl. Phys. B (Proc. Suppl.)}}
\def \NP {{ Nucl. Phys.}}
\def \RMP {{ Rev. Mod. Phys.}}
\def \JGP {{ J. Geom. Phys.}}
\def \CQG {{ Class. Quant. Grav.}}
\def \MPL {{Mod. Phys. Lett.}}
\def \IJMP {{ Int. J. Mod. Phys.}}
\def \JHEP {{ JHEP}}
\def \PR {{Phys. Rev.}}
\def \JMP {{J. Math. Phys.}}
\def \GRG{{Gen. Rel. Grav.}}
\begin{titlepage}
\null\vspace{-62pt} \pagestyle{empty}
\begin{center}
\rightline{AEI-2008-045}
\rightline{CCNY-HEP-08/4}
\rightline{July 2008}
\vspace{.6truein} {\Large\bfseries
The Hamiltonian Analysis for Yang-Mills Theory on ${\mathbb R}\times S^2$}\\
\vskip .1in
\vspace{.6in}
{\large ABHISHEK AGARWAL$^a$}  and {\large V.P. NAIR$^b$}\\
\vskip .2in
{\itshape $^a$Max Planck Institut f\"{u}r Gravitationsphysik\\
Albert Einstein Institut\\
Am M\"{u}hlenberg-1, D14476, Potsdam, Germany}\\
\vskip .1in
{\itshape $^b$Physics Department\\
City College of the CUNY\\
New York, NY 10031}\\
\vskip .1in
\begin{tabular}{r l}
E-mail:&{\fontfamily{cmtt}\fontsize{11pt}{15pt}\selectfont abhishek@aei.mpg.de}\\
&{\fontfamily{cmtt}\fontsize{11pt}{15pt}\selectfont vpn@sci.ccny.cuny.edu}
\end{tabular}

\fontfamily{cmr}\fontsize{11pt}{15pt}\selectfont
\vspace{.8in}
\centerline{\large\bf Abstract}
\end{center}
Pure Yang-Mills theory on ${\mathbb R} \times S^2$ is analyzed in a gauge-invariant Hamiltonian formalism. Using a suitable coordinatization for the sphere and a gauge-invariant matrix parametrization for the gauge potentials, we develop the Hamiltonian formalism in a manner that closely parallels previous analysis on ${\mathbb R}^3$.
The volume measure on the physical configuration space of the gauge theory,
the nonperturbative mass-gap and the leading term of the vacuum wave functional are discussed
using a point-splitting regularization.
 All the results carry over smoothly to known results on ${\mathbb R}^3$ in the limit in which the sphere is de-compactified to a plane.
\end{titlepage}
\pagestyle{plain} \setcounter{page}{2}

\section{Introduction}
This paper will set up the framework for a Hamiltonian analysis of Yang-Mills theories in 2+1 dimensions, for the case of the spatial manifold being a two-sphere $S^2$ of finite radius
$r$.
Specifically, we formulate the  theory in a gauge-invariant Hamiltonian framework by suitably coordinatizing the sphere and utilizing the matrix parametrization of the gauge potentials.
The invariant volume measure on the physical configuration space of Yang-Mills theory on ${\mathbb R} \times S^2$ and the computation of the mass gap and vacuum wave functional are discussed.
Even though, many of the technical details are rather different from that of the gauge theory on ${\mathbb R} \times {\mathbb R}^2$, the final results final results that we obtain are in complete agreement with the expectations from the behavior of the gauge theory when the spatial manifold is a plane.

The Hamiltonian approach to Yang-Mills theories in 2+1 dimensions, developed by Karabali, Kim and Nair (KKN) \cite{KKN1}, has the potential to address a number of nonperturbative questions such as confinement, mass gap and screening \cite{KKNVWF}-\cite{Abe}. The initial calculations led to a formula for the string tension which has been shown to be in good agreement with lattice calculations \cite{teper-1}. The possibility of  incorporating  glueballs within this scheme has been explored in a number of papers \cite{glueballs1}. More recently, the screening of the adjoint and other screenable representations has been discussed \cite{AKN}.
The seminal idea for all these calculations has been a matrix parametrization for the gauge potentials which makes the implementation of gauge invariance particularly simple.
After transforming the Hamiltonian to this parametrization,  the vacuum wave function was calculated in a $1/e^2$-expansion, where $e$ is the Yang-Mills coupling constant.
This is something like a continuum strong coupling expansion but, in this context, it is important to keep in mind that there is really no suitable expansion parameter
for the Yang-Mills theory in 2+1 dimensions (except possibly for $1/N$).
The role of the coupling constant $e$ is simply that modes of the field with momenta $k\ll e^2$
should be treated nonperturbatively, while modes with momenta $k\gg e^2$ can be treated in perturbation theory.
For elucidating the nonperturbative structure of the theory, an expansion in terms of
$k/e^2$ can be suitable, although corrections need to be worked out carefully.
One may ask whether the $1/e^2$-expansion can be phrased in terms of a dimensionless
fixed parameter. Such a  characterization will need another dimensional fixed parameter
in the theory. Finite volume for the spatial manifold can provide such a parameter.
This is one of the motivations for considering $S^2$. The continuum strong coupling
expansion can then be considered as an expansion in powers of
$1/(e^2r)$.

More generally, the analysis  of the theory on manifolds of the form ${\mathbb R}\times \Sigma$, where $\Sigma$ is a Riemann surface can be very useful. The case of a torus, for instance, can be related to the theory at finite temperature and so many features related to deconfinement can be analyzed\footnote{For some recent work in this direction see \cite{Abe}}. One may regard the present work, for ${\mathbb R}\times S^2$ as the zeroth step in a more general analysis.

Apart from the motivations outlined above, another reason to explore the KKN formalism on various nontrivial spatial geometries is the following. The formulae for the mass-gap, string-tension and vacuum wave functional of the gauge theory on ${\mathbb R} \times {\mathbb R} ^2$ , obtained first by explicit computations, provide us with some insights into the  geometric features that tie these quantities together. It was recently argued in \cite{robustness}  that these quantities are related to each other by some rather generic features such as Lorentz invariance and the two dimensional anomaly computation which ultimately dictates the measure on the configuration space of the gauge theory. By computing mass-gap and vacuum wave function for the strongly coupled theory on ${\mathbb R} \times S^2$ we provide further  evidence in favor of these arguments.

There is yet another reason why the analysis of Yang-Mills theory on ${\mathbb R}\times S^2$ can be interesting.
In the case of this theory with additional matter degrees of freedom corresponding to
sixteen supercharges, there is a recent proposal
about the gravity dual description.
The computation of  the masses of operators built out of
the scalars fields in the theory has been carried out, from the string theory side, in
the leading strong
coupling limit. The analysis developed here, when augmented by the addition of matter fields, has the potential for the computation of the same quantities from the gauge theory side.
This can, obviously, be useful in elucidating the
gauge-gravity duality.

The paper is organized as follows. In the next section we construct the gauge-invariant variables appropriate for the Hamiltonian analysis on $S^2$.
 The volume measure on the configuration space, obtained as the Jacobian for this change of variables is reported in the next section; details of the derivation of the measure are presented in the first appendix. In sections 4 and 5, we provide expressions for the Hamiltonian in terms of the gauge-invariant quantities. As with the analysis on the plane, special attention needs to paid to the regularization of the Hamiltonian and various other local composite operators. A point-split regularization scheme that is compatible both with the symmetries of the sphere and the `holomorphic' invariance of the gauge theory is also elaborated upon in these sections. In section 6 we use the formalism to compute the leading order term
 (the term with two powers of the current) in the vacuum wave functional  of the theory. Many of the technical details relevant to this computation, such as an expression for the Hamiltonian in momentum space, are contained in the second appendix.

 \section{The matrix parametrization}

As is standard in Hamiltonian analyses, we shall use the $A_0=0$ gauge.
We then have the two spatial components $A_z, A_\bz$\footnote{The convention here is
that $z = x_1-ix_2, \hspace{.2cm} \bar{z} = x_1 + ix_2$ while $A_z = \frac{1}{2}(A_1+iA_2), \hspace{.2cm} A_\bz = \frac{1}{2}(A_1 - iA_2).$}. For ${\mathbb R}^2$,
we used the parametrization
\beq
A_z = - \del_z M ~M^{-1}, \hskip .3in A_\bz = M^{\dagger -1} \bdel M^\dagger
\label{sph1}
\eeq
We want to construct the analogue of this for $S^2$. For this, it is convenient to think of
$S^2$ as $SU(2)/U(1)$ and use the group translation operators (i.e., angular momentum operators) as covariant derivatives. Such an analysis (which can be extended to certain higher dimensional spaces as well) was used extensively in the study of the quantum hall effect (and its non-commutative analog) in \cite{QH-1}. Using this coordinatization in the present context allows us to follow the analysis carried out on the plane extremely closely.

We can use a group element $g\in SU(2)$ to parameterize the two-sphere.
Explicitly, the standard complex coordinates on $S^2$ may be related to $g$ via the parametrization
\beq
g = {1\over \sqrt{1+z\bz}} \left[ \begin{matrix}
1 ~~& ~~z\\
-\bz &~~1\\
\end{matrix}
\right] ~\left( \begin{matrix}
e^{i\psi/2}&0\\
0&e^{-i\psi/2}\\
\end{matrix}
\right)\label{sph2}
\eeq
In terms of this coordinatization, the non-vanishing components of the metric tensor
and the volume element on the sphere are given by
 \beq g_{z\bz} = g_{\bz z} =
\frac{r^2}{\pi (1+z\bz)^2}, \hskip .3in
d\mu = \frac{r^2 dzd\bar{z}}{\pi(1+z
\bz )^2}
\label{sph3}
\eeq
In our conventions, the area of the sphere is $r^2$
\footnote{It is also understood that $dzd\bar{z}$ is only a short-hand notation for $\frac{1}{2i}dzd\bar{z} = d^2x$}.
The volume element on the sphere is the the natural one induced from
the measure on $SU(2)$, when the volume of the Lie group is
normalized to $r^2$.
In terms of the usual angular coordinates of $S^2$,
\beq
z = \tan(\theta /2)e^{i\phi}\label{sph4}
\eeq
Functions which are well defined on the sphere are independent of the $U(1)$ angle
$\psi$. We may think of them as $U(1)$-invariant functions on the group $SU(2)$.
For such functions, we may also write
\beq
\int d\mu (S^2) ~f(z, \bz)  = \int  \frac{r^2 dzd\bar{z}}{\pi(1+z \bz
)^2}~f(z,\bar{z})= \int {d\mu(g)\over 2\pi}~ f(g)
\label{sph5}
\eeq
where $d\mu (g)$ is the Haar measure for the group $SU(2)$.

Consider the left and right translations operators on $SU(2)$ defined by
\beq
L_a ~g = \half \sigma_a ~g, \hskip .4in R_a ~g = g~\half \sigma_a
\label{sph6}
\eeq
where $\half \sigma_a$ are the generators of $SU(2)$ in the $2\times 2$ matrix representation.
We will use $R_\pm = R_1 \pm i R_2$ as the translation operators on
$S^2 = SU(2)/U(1)$, with $R_3$ as the $U(1)$ generator. Thus functions on
$S^2$ are invariant under $R_3$. We define the Wigner $\D^j_{mn}(g)$-functions
as the representative of $g$ in the spin $j$-representation,
\beq
\D^j_{mn}(g) = \la j, m\vert ~{\hat g} ~\vert j, n\ra
\label{sph7}
\eeq
Functions on $S^2$ can be expanded in terms of $\D^j_{m0}(g)$.

Corresponding to $R_\pm$, we introduce the gauge potentials
$A_\pm$, setting $A_0=0$ as on the plane.
The matrix parametrization of the fields may then be written as
\beq
A_+ = -\frac{\sqrt{\pi}}{r}(R_+M)M^{-1}, \hskip .3in A_- = \frac{\sqrt{\pi}}{r}(M^{\dagger -1}R_-
M^{\dagger})\label{sph8}
\eeq
It is instructive to compare these with the components in the coordinate basis.
For functions on $S^2$,
\beq
R_+ = (1+z\bz)~\partial_z, \hskip .3in R_- =
(1+z\bz)~\partial_{\bz}\label{sph9}
\eeq
Correspondingly, the coordinate components $A_z, A_\bz$ of the
gauge potentials are given by
\beq
A_+ = \sqrt{\pi}(1+z\bz)~A_z, \hskip .3in A_- = \sqrt{\pi}(1+z\bz)~A_{\bz}\label{sph10}
\eeq
With $z = (x_1 -i x_2 )/r$, we see that the large $r$ limit returns the
parametrization of the fields on the
plane. $A_\pm$ are the components of the potentials in the tangent frame basis.
Gauge transformations on the sphere are given by
\beq
A_{\pm} \rightarrow UA_{\pm}U^{-1} -\frac{\sqrt{\pi}}{r} (R_{\pm}U)U^{-1}\label{sph11}
\eeq
These transformations obviously are equivalent to left translations of $M$ by $U$,
$M \rightarrow M^U(x) = U(x) M(x)$.

As on the plane, the matrix parametrization results in a new gauge symmetry, the so-called holomorphic invariance. From the definitions (\ref{sph8}), it is easy to see that $M$ and $M\bar{V}(\bar{z})$, where $\bar{V}(\bar{z})$ is a matrix that depends only on the anti-holomorphic coordinate $\bar{z}$, result in the same gauge potentials.
While there are no globally defined antiholomorphic functions on the sphere, it is necessary to be able to use the parametrization (\ref{sph8}) patchwise on the spatial manifold and hence it is important to keep track of this symmetry.
In particular, this symmetry needs to be preserved in any physically meaningful computation that one might intend to carry out in the theory. In particular regularization schemes must preserve this symmetry. We use this as a guiding principle to do various regularized computations later in the paper.

Starting with the Yang-Mills action, we can now obtain the canonical one-form
as
\beqar
\Theta &=& -4\int d\mu ~\pi(1+z\bz )^2~\mbox{Tr}(E_z\delta A_{\bz}
+ E_{\bz} \delta A_z)\nonumber\\
&=& -4\int d\mu ~\mbox{Tr}(E_+\delta A_{-} +
E_{-} \delta A_+)
\label{sph12}
\eeqar
As for $A$, the tangent frame components of $E$ are related to the coordinate components by $E_+ = \sqrt{\pi}(1+z\bz)E_z,~E_- = \sqrt{\pi}(1+z\bz)E_{\bz}$.

Upon using the parametrization (\ref{sph8}), $\Theta$ becomes
\beqar
\Theta &=& 2\int d\mu ~\mbox{Tr}(\bp~ \delta M^{\dagger } M^{\dagger
-1} + p~M^{-1}\delta M)\nonumber\\
p &\equiv& p^a t^2 =
-2\frac{\sqrt{\pi}}{r}\rpl(M^{-1}E_- M)\label{sph13}\\
\bp &\equiv& \bp ^at^a =
2\frac{\sqrt{\pi}}{r}\rmin(M^{\dagger}E_+M^{\dagger -1})\nonumber
\eeqar
Here $\rpl, \rmin$ denote the translation operators
$R_+, R_-$ acting on vectors.
Recall that $R_+, R_-$ correspond to the Levi-Civita covariant derivatives; their action on vectors will be different from their action on scalar functions because of the Levi-Civita or
spin connection.
The commutation rule $[R_+, R_- ]= 2 R_3$ can be interpreted as the commutator of two covariant derivatives, with $R_3$ being proportional to the Riemann tensor of $S^2$
multiplied by the spin operator of the tensor on which it acts.
Explicitly,
\beq
\rpl =
(1+z\bz)\partial_z -\bz, \hskip .3in\rmin = (1+z\bz)\partial_{\bz} - z\label{sph14}
\eeq

The Green's functions for the operators $\rpl, \rmin$ can now be defined
as follows.
\beqar
\rpl (z) G_+(z,w) &=& \delsp (z-w) -\frac{1}{r^2}\nonumber\\
\rmin (z) G_-(z,w) &=& \delsp
(z-w) -\frac{1}{r^2}\nonumber\\
G_+(z,w) = \frac{1}{r^2}\frac{1+\bar{w}z}{\bz - \bar{w}}, &&\hskip .2in
G_-(z,w) = \frac{1}{r^2}\frac{1+w\bz}{z - w} \label{sph15}
\eeqar
Here
\beq
\delsp(z-w) = \frac{\pi}{r^2}
(1+z\bz )^2\delta ^{(2)}(z-w) \label{sph16}
\eeq
is the Dirac delta function on
the two-sphere appropriate to the tangent frame we are using.
 In the above formulae for the Green's functions, the
subtraction of $\frac{1}{r^2}$, has to do with the existence of the (constant)
zero mode for the Laplace operator on the two-sphere. This is to be
contrasted with the situation one encounters for ${\mathbb R}^2$, in which
case there is
no such zero mode to be subtracted. As is evident, the zero mode contribution goes to zero as $r\rightarrow \infty$.

Using these Green's functions, the expressions  (\ref{sph13}) can be inverted,
with the
the electric field components given in terms of the gauge-invariant
momentum operators $p, \bp$ as
\beqar
 E_+^k(x) &=& \frac{ir}{2\sqrt{\pi }}(\mathcal{M}^\dagger)^{ak}(x)\int_y d\mu_y~G_-(x,y)~\bar p^a(y)\nonumber \\
E_-^k(x) &=& -\frac{ir}{2\sqrt{\pi }}(\mathcal{M})^{ka}(x)\int_y
d\mu_y~G_+(x,y)~p^a(y)\label{sph17}
\eeqar
Here $\mathcal{M}$ is the adjoint
representative of the matrix $M$ i.e.,
 \beq
 \mathcal{M}^{ab} =
2\mbox{tr}(t^aMt^bM^{-1})\label{sph18}
\eeq
Since the Green's functions involve
subtractions of a zero mode, we notice that the equivalence of
(\ref{sph13}) and (\ref{sph17}) require that
\beq
\int d\mu_x~ p^a(x) = \int
d\mu_x~ \bar p^a(x) = 0 \label{sph19}
\eeq
This is equivalent to the statement that the total charge on the closed manifold $S^2$ must be zero. Notice that
$p,\bar{p}$ are expressible as total derivatives of the electric field components and so
the integrals correspond to the total charge on the space.
(If the definition of $p, \bp$ in (\ref{sph13}) is integrated on a space with a boundary, we would get the integrals of the electric field components over the boundary, which would be the charge.)
Another way to phrase this is by
observing, again  from their definition (\ref{sph13}), that the operators $p,\bp$ have no $j=0$ mode, hence their volume
integral must vanish.

The canonical one-form (\ref{sph13}) allows us to read off the basic
commutation relations as
\beqar
 [p^a(x),p^b(y)] &=&~~ f^{abc}p^c(x)~\delsp (x-y)\nonumber\\
{} [\bar{p}^a(x),\bar{p}^b(y)] &=& -f^{abc}\bar{p}^c(x)~\delsp (x-y)\nonumber\\
{}[p^a(x), M(y)] &=& -iM(y)t^a~ \delsp (x-y) \label{sph20}\\
{}[\bar{p}^a(x), M^{\dagger}(y)] &=& -it^a M^{\dagger}(y)~\delsp (x-y)\nonumber
\eeqar
All the other commutators vanish.
\section{The measure of integration}

The evaluation of the measure of integration for the gauge fields, which determines the inner product for wave functionals, is the next step. We will work this out in the appendix, but
the result is basically unchanged from the result on the plane.
On the plane, we get the WZW action for the gauge-invariant combination
$H = M^\dagger M$. This action involves the kinetic term and the WZ term.
The latter is a differential form and hence does not depend on the metric; it will
therefore have the same form on the plane and the sphere. (Recall that
the result on the plane is also for the case
with the point at infinity added, so that it is also topologically
a sphere, although metrically distinct.) The kinetic term is again essentially the same,
because the sphere metric is conformally flat and the kinetic term is classically conformally invariant. Thus the result can be easily written down as
\beqar
\det (- D_+ D_- ) &=& {\rm constant}~ \exp (2 c_A S_{WZW})\nonumber\\
S_{WZW}(H) &=& {1\over 2} \int d\mu~\Tr  \left[ R_+ H~ R_- H^{-1}\right]\nonumber\\
&&\hskip .2in+{i\over 12\pi} \int d^3x~ \epsilon^{\mu\nu\alpha}
~\Tr \left[ H^{-1}\del_\mu H~H^{-1}\del_\nu H~H^{-1}\del_\alpha H\right]
\label{sph21}
\eeqar
where $c_A$ is the adjoint Casimir defined by $c_A \delta_{ab} = f_{amn}f_{bmn}$;
$c_A =N$ for $SU(N)$.

\section{The Hamiltonian}

\subsection{${\cal H}$ in terms of gauge-invariant variables}

The next step in setting up the formalism is the construction of the Hamiltonian
in terms of the gauge-invariant variables. Since these involve functional operators,
regularization is important. We shall start with the naive expressions here and
discuss in the next subsection how the regularization can be included.

The Hamiltonian operator is ${\cal H} = T + V$, where the kinetic energy operator
$T$ is
\beqar
 T &=& 2 e^2\int d\mu~ E_+^a~E_-^a \nonumber \\
 &=& \frac{ e^2r^2}{2\pi }\int d\mu(x) \int [d\mu(y) d\mu(w)]G_-(x,y)\bar{p}^a(y)K^{ab}(x)G_+(x,w)p^b(w)
 \label{sph22}
 \eeqar
 where
 \beq
  K^{ab} = 2\mbox{Tr}(t^aHt^bH^{-1}) = \left(\mathcal{M}^{\dagger}\mathcal{M}\right)^{ab}
  \label{sph23}
  \eeq
  $K^{ab}$ is the adjoint representation of $H$. In the second line
  of (\ref{sph22}), we have used the expressions (\ref{sph17}) for the $E_\pm$.

  The potential energy, written out in terms of the $R_\pm$derivatives, is
 \beq
V =  \frac{4}{\pi e^2}\int d\mu\mbox{Tr}\left(\frac{\sqrt{\pi }}{r}(\rpl A_- - \rmin A_+)
+ [A_+,A_-]\right)^2\label{sph24}
 \eeq
 Notice that the parametrization of the potentials in (\ref{sph8}) can be expressed as
 \beqar
 A_+ &=& M^{\dagger -1} \left[ - {\sqrt{\pi} \over r} R_+ H ~H^{-1}\right] +
 {\sqrt{\pi} \over r} M^{\dagger -1} R_+ M^\dagger\nonumber\\
  A_- &=& M^{\dagger -1} \left[ ~0~\right] ~M^\dagger +
 {\sqrt{\pi} \over r} M^{\dagger -1} R_- M^\dagger
 \label{sph25}
 \eeqar
 In other words, the potentials $(A_+, A_-)$ are the gauge transform of $(-\sqrt{\pi} R_+H H^{-1}/r, 0)$
 by the complex matrix $M^\dagger$.
 It is then easy to see that the potential energy can be written as
 \beq
V = \frac{2\pi ^3}{e^2N^2}\int d\mu \left( \frac{\rmin J^a}{r} \frac{\rmin J^a}{r}\right)\label{sph26}
\eeq
where the current $J^a$, as in the  case of Yang-Mills on ${\mathbb R}\times {\mathbb R}^2$, is
\beq
J^a = \frac{2N}{\pi r}\mbox{Tr}(t^a
R_+HH^{-1})\label{sph27}
\eeq

As the radius of the sphere $r$ becomes large, the sphere is well approximated by
the plane. It is
interesting to see how the expressions for various quantities on the sphere
go over to the corresponding quantities on the plane, as described in \cite{KKN1}. The coordinates on the plane $w, \bw$ are related to $z, \bz$ by
 $z,\bar{z} = \frac{w}{r}, \frac{\bar{w}}{r}$.
It is easily verified that, as $r\rightarrow  \infty$, $d\mu \rightarrow d^2x/\pi$. Similarly,
$R_\pm /r \rightarrow \del, \bdel$, $A_\pm \rightarrow \sqrt{\pi} A, \sqrt{\pi} \bA$.
The current $J^a$ goes over to its planar image without any additional factors.

\subsection{Regularization}

We now turn to the question of regularization.
In the case of Yang-Mills on ${\mathbb R} \times {\mathbb R}^2$,
a point-splitting regulator consistent with the holomorphic invariance was used.
The basic ingredient necessary for this was a `smoothed
out' version of the Dirac delta function on the plane. Specifically, the choice was
 \beq
 \sigma_p(z,w,\epsilon) =
{1\over \pi \epsilon} \exp\left(- {\vert z - w\vert^2 \over \epsilon}\right)\label{sig1}
\eeq
Here
$z$ and $w$ are
complex coordinates on the plane e.g. $z = (x_1-ix_2)$. This expression
for the regularized $\delta$-function reduces to the planar $\delta$-function as $\epsilon
\rightarrow 0$. Thus $\epsilon $ can be considered as the ultraviolet cutoff
for the theory;
of course, all computations are to be done with finite $\epsilon$ which is
allowed to approach zero only after physical quantities are computed.

The first step towards regularizing the theory on the sphere is
the sphere-analogue of the above expression for $\sigma_p$.
Since $S^2= SU(2)/U(1)$, functions on the sphere can be
expressed in terms of the Wigner $\mD$-functions
$\mD ^l_{m,0}$ which are proportional to the usual spherical
harmonics $Y_{l,m}$.
Denoting an arbitrary element of $SU(2)$ by $g$,
any linear combination
\beq
f(g) = \sum_{l,m} a_{lm}\mD^l_{m,0}(g) \label{sig2}
\eeq
is a function on $S^2$.
The Wigner functions are
normalized as
\beq \int d\mu(g) \mD^{l*}_{m,0}(g)\mD^{l'}_{n,0}(g)
= \frac{r^2 \delta ^{l,l'}\delta _{m,n}}{2l+1} \label{sig3}
\eeq
where $d\mu(g)$ is the invariant measure (the Haar measure) on $SU(2)$.
In terms of the local coordinates $z, \bz$ for the sphere,
it can be given explicitly as
\beq
d\mu (g) = {d\theta_2 \over 2\pi}~
\frac{r^2 dzd\bz}{\pi(1+z\bz)^2} \label{sig4}
\eeq
where $\theta_2$, with $0 \leq \theta_2 \leq 2\pi$, is the extra $U(1)$ angle.
The mode decomposition of
of the delta function is thus given by
\beq
\delta(g,g') =
\sum_{l,m}\frac{(2l+1)}{r^2}\mD ^l_{m,0}(g) \mD^{l*}_{m,0}(g') = \sum_l
\frac{(2l+1)}{r^2}\mD^{l}_{0,0}(g'^{\dagger}g).\label{sig5}
\eeq

We need
a one-parameter family of functions which are consistent with the
coset space properties of the sphere and which reduce to the
expression above when the parameter goes to zero. Before working out such an extension,
it is useful to recall
some relations between the coset space
representations of points on $S^2$ and their usual polar angle
depictions.

An arbitrary $SU(2)$ element can be parameterized as
\beq
g =
\symmat{g_{11}}{g_{12}}{g_{21}}{g_{22}} =
\symmat{u_2^*}{u_1}{-u_1^*}{u_2}, \hskip .2in  \det~g = |u_1|^2+|u_2|^2 =
1\label{sig6}
\eeq
As is well known, the two-sphere will require at least two coordinate patches,
one on the northern hemisphere (which can be extended to everywhere on $S^2$ except
at the south pole) and the other on the southern hemisphere.
In terms of $u_1, u_2$, one patch has $u_2\neq 0$ and other has $u_1\neq 0$.
On the first one, for example, we can define the local coordinate
$z= u_1/u_2$. In this case, the general $SU(2)$ element
(\ref{sig6}) can be brought to the form
\beq
g =
\frac{1}{\sqrt{1+z\bz }}\symmat{1}{z}{-\bz}{1}\symmat{e^{-i\theta _2
}}{0}{0}{e^{i\theta_2}}
\label{sig7}
\eeq
where $\theta_2$ is the argument of $u_2$.
The change of variables $z = \tan({\theta }/{2})e^{i\phi}$ will bring us to the standard
parametrization of the sphere in terms of the polar coordinates $\theta, \phi $.

The angle $\Omega $ between the unit vectors in the directions
$(\hat{\theta},
\hat{\phi })$ and $(\hat{\theta'}, \hat{\phi '})$ is given by
\beqar
\cos(\Omega)  &=& \left(\cos(\theta) \cos(\theta ') + \sin(\theta)
\sin(\theta ')\cos(\phi - \phi')\right) \nonumber\\
&=&  \frac{(1+\bz
w)(1+\bar{w}z) - |z-w|^2}{(1+z\bz)(1+w\bar{w})} \label{sig8}
\eeqar
In terms of group parameters, we may write
\beq
\cos(\Omega) = U_{11}U_{22}+ U_{21}U_{12} \label{sig9}
\eeq
where
\beq
U =
g'^\dagger g = \frac{1}{\sqrt{(1+z\bz)(1+w\bar{w})}}\symmat{1+\bz
w}{z-w}{-\bz + \bar{w}}{1+\bar{w}z}\label{sig10}
\eeq
The geodesic distance two points is given by $r \Omega$; we can also relate the angle
$\Omega$ to
the chordal distance $4 \Delta (z,w)$ by $ 1- \cos (\Omega )= 2 \Delta (z,w)$, where
\beq
\Delta(z,w) =
\frac{|z-w|^2}{(1+z\bz)(1+w\bar{w})}\label{sig11}
\eeq
With these formulae and the standard expression for the spherical harmonics, we see that
$\mD_{0,0}$ is given in terms of the Legendre polynomials $P_l$ by
\beq
\mD^l_{0,0}(g'^\dagger
g) = P_l(\cos (\Omega))\label{sig12}
\eeq
with $g,g'$ are the group
elements corresponding to $z,w$, respectively.

We now consider the function $e^{t\cos(\Omega)}$ which can be
expressed using the Gegenbauer expansion formula as
\beq
e^{t\cos(\Omega)} =
\left(\frac{2}{t}\right)^{1/2}\Gamma(1/2)\sum_{l=0}^{\infty}
(l+\frac{1}{2})P_l(\cos(\Omega ))I_{l+\frac{1}{2}}(t)\label{sig13}
\eeq
Here $I_\nu$
is a modified Bessel function of order $\nu$.
Writing $2t
= {1/\epsilon}$, we
see immediately that, for small $\epsilon$ (large $t$),
\beq
{1\over r^2\epsilon}{\exp\left(-\frac{1}{2\epsilon}(1-\cos(\Omega)\right)}
~{\approx} ~\sum_{l=0}^{\infty}
e^{-l(l+1)\epsilon }\frac{(2l+1)}{r^2}P_l(\cos(\Omega))
\label{sigsp}
\eeq
The large $t$-asymptotic formula for the modified Bessel function has been used
for this simplification,
\beq
I_{\nu }(t) \stackrel{\mbox{large t}}{\approx} \frac{1}{\sqrt{2\pi
t}}\exp\left(t-\frac{\nu ^2-\frac{1}{4}}{2t}\right)\label{sig14}
\eeq

Notice that the right hand side of
(\ref{sigsp})  is nothing but the heat kernel on the two sphere since
 \beq
(\partial_\epsilon - \mathcal{R}_-R_+)\sum_{l=0}^{\infty}
e^{-l(l+1)\epsilon }(2l+1)P_l(\cos(\Omega))=0 \label{sig15}
\eeq
From the
asymptotic expansion (\ref{sigsp}) and the formula (\ref{sig5})
for $\delta (g, g')$, we see that
\beq
Lim_{\epsilon\rightarrow 0} ~{1\over{r^2\epsilon}}~ {\exp\left(-\frac{1}{2\epsilon}(1-\cos(\Omega)\right)} =
\delta(g,g')\label{sig16}
\eeq
Thus a sharply peaked Gaussian function on the sphere can be
expressed as
\beqar
\sigma(z,w,\epsilon) &=&
{1\over{r^2\epsilon} } {\exp\left(-\frac{1}{2\epsilon}(1-\cos(\Omega)\right)}=
\frac{e^{-\Delta(z,w)/\epsilon }}{r^2\epsilon}\nonumber\\
&\rightarrow&\delta (g,g'), \hskip .2in {\rm as}~ \epsilon \rightarrow 0\label{sig17}
\eeqar

Using (\ref{sig11}) $\sigma$ can also be expressed
manifestly in terms of the $z,w$ variables as
\beq
\sigma(z,w,\epsilon) = {1\over {r^2\epsilon}}~{\exp\left(-\frac{|z-w|^2}{\epsilon
(1+z\bz)(1+w\bar{w})}\right)}\label{sig18}
\eeq
\subsection{Regularized Expression for the Kinetic Energy Operator}

The regularization of the kinetic energy operator was carried out in
\cite{KKN1} at the level of the `momentum' operators
$p^a, \bp^a$. With the definition of $\sigma (z, w, \epsilon )$ given above,
we can follow the same procedure on the sphere,
defining the regularized operators by
\beqar
p^a \rightarrow \int d\mu(y)\sigma(x,y,\epsilon) (K^{-1}(y,\bar{x})K(y,\bar{y}))^{ab}p^b(y)\nonumber \\
\bp ^a \rightarrow \int d\mu(y)\sigma(x,y,\epsilon) (K(x,\bar{y})K^{-1}(y,\bar{y}))^{ab}\bp^b(y)
\label{KE1}
\eeqar
The regularized expressions have the same transformation properties under holomorphic transformations and reduce to the unregulated expressions if $\epsilon $ is let go to zero.
The parameter $\epsilon $ serves has a short distance cut-off.

The regularized expression for the kinetic energy operator is can now be given as
\beq
T = \frac{r^2 e^2}{2\pi }\int d\mu(u)d\mu(v) ~\Pi^{rs}(u,v)~\bp^r(u)~p^s(v)
\eeq
where
\beq
\Pi^{r,s}(u,v) = \int d\mu(x) \left( \mg^{ar}_-(x,u)\right) K^{ab}(x)\left(\mg ^{bs}_+(x,v)\right)
\eeq
The regularized Green's functions occuring in this formula
are given by
\beqar
 \mg_+^{ab}(x,y) &=& \int d\mu(u) G_+(x,u)\sigma (u,y,\epsilon)(K^{-1}(y,\bar{u})K(y,\bar{y})^{ab}\nonumber \\
\mg_-^{ab}(x,y) &=& \int d\mu(u) G_-(x,u)\sigma (u,y,\epsilon)(K(u,\bar{y})K^{-1}(y,\bar{y})^{ab}.
\label{reggf}
\eeqar
\section{The Hamiltonian in terms of currents}

We shall start with the expression of the kinetic energy in terms of currents.
As in the case of the theory on the plane, the wave function can be taken to be a function of the current defined in (\ref{sph27}).
Since the kinetic energy operator is quadratic in the $p,\bar{p}$ variables, it is easily seen from the chain rule for functional differentiation that $T$ will contain a term with one derivative with respect to $J^a$ and another term with two derivatives with respect to $J^a$.
The coefficients of these terms can be found by evaluating the action of $T$ on
functional involving, at most, two powers of the current.
The commutation relations we shall need  for this calculation are
 \beqar
  [p^s(v), J^a(z)] &=& -\frac{iN}{r\pi}K^{as}(z)R_{+z}\delta^2_{S^2}(v-z)\nonumber\\
  {} [\bp^r(u), J^a(z)] &=& -i\mD ^{br}_z \delta ^2_{S^2}(z-u)\label{kej1}
  \eeqar
where $\mD$ is the holomorphic covariant derivative given by
\beq
\mD^{ar}_z= \frac{N}{r\pi}R_{+z}\delta ^{ar} + i f^{arc} J^c(z)\label{kej2}
\eeq
Other commutation relations which are useful for this calculation are
\beqar
 [p^a(x), K^{mn}(y)] &=& f^{anc}K^{mc}\delta^{(2)}_{S^2}(x-y)\nonumber\\
{} [\bp^a(x), K^{mn}(y)] &=& f^{mac}K^{cn}\delta^{(2)}_{S^2}(x-y)\label{kej3}
\eeqar

The action of $T$ on $J^a$ can now be simplified as
\beqar
 T~ \int_z c^a(z)J^a(z) &=& -\frac{iNr e^2}{2\pi^2}\int_{z,u,v,x}\mg_-^{mr}(x,u)K^{ms}(x)G_+(x,v)~[\bp^r(u),K^{as}(z)]\nonumber\\
 &&\hskip .7in \times (R_{+z}\delta^{(2)}_{S^2}(v-z))c^a(z)\label{kej4}
 \eeqar
We have left, $G_+$ in the unregulated form in this expression; this is adequate for this calculation.
The right hand side of (\ref{kej4}) can be simplified by noting that
\beqar
 \int_{z,v} G_+(x,v) K^{as}(z)[R_+(z)\delta^2_{S^2}(v-z)]c^a(z) &=& \int_z c^a(z)K^{as}(z)[R_+(z)G_+(x,z)]\nonumber\\
 &=& -c^a(x)K^{as}(x)\label{kej5}
 \eeqar
In the above manipulations it is useful to recall that
$G_+ (x,z)$ is the Green's function for $R_{+z}$, i.e.,
\beq R_+(z)G_+(x,z) = (1+z\bz)\partial_z \left(\frac{1+x\bz}{r^2(\bar{x}-\bz)}\right) = -\delta^2_{S^2}(z-x)\label{kej6}
\eeq
Using this in (\ref{kej4}) we have,
\beq
 T ~ \int_z c^a(z)J^a(z) = \frac{iNre^2}{2\pi^2}f^{arl}\int_z \mg_-^{lr}(z,z)c^a(z)
 \label{kej7}
 \eeq
The coincident point limit of the Green's function $ \mg_-^{lr}(z,z)$ can be obtained by expanding the definition (\ref{reggf}) around $u=x$.
This leads to
\beqar
\mg_- ^{ab}(x,x) &=& \delta ^{ab}\int d\mu(u) G_-(x,u)\sigma(u,x,\epsilon) \nonumber\\
&& \hskip .2in+ ((\partial K)K^{-1})^{ab}(x)\int d\mu(u) G_-(x,u)(u-x)\sigma(u,x,\epsilon) + \cdots \nonumber \\
& \equiv& \delta ^{ab}I_1 + I_2 + \cdots \label{coincident}
\eeqar
The higher terms which are not shown here are
of at least $\mathcal{O}(\epsilon )$ and are irrelevant for this calculation.
The contribution of the term involving $I_1$ in (\ref{coincident}) to (\ref{kej7})
is zero since the trace over the color indices vanishes.
The integrand in $I_2$ has no singularities and it can be evaluated by taking $\epsilon \rightarrow 0$ to get $r^2 I_2 = (R_+K)K^{-1}$, so that
\beq
\mg_-^{lr}(z,z) = -\frac{1}{r^2}[(R_+K)K^{-1}]^{lr}(z) = \frac{i\pi}{Nr}f^{lrc}J^c(z)\label{kej8}
\eeq
Equation (\ref{kej7}) can now be simplified as
\beq
T~ \int_z c^a(z)J^a(z) = \frac{ e^2N}{2\pi} \int_z c^a(z)J^a(z)\label{kej9}
\eeq
Notice that the parameter for the mass gap is the same as on ${\mathbb R} \times {\mathbb R}^2$; of course, this is not surprising, since it arises from the two-dimensional anomaly, as explained elsewhere.

The calculation given above shows that the term in $T$ involving one derivative with respect to $J$ can be written as
\beq
T_1 =  \frac{ e^2N}{2\pi } \int_z J^a(z)\frac{\delta}{\delta J^a(z)}
\label{t1}
\eeq
It may be worth pointing out that
$\frac{\delta}{\delta J^a(z)}$ is only short-hand notation for the operator whose
commutation relation is given by
\beq
\left[\frac{\delta}{\delta J^a(z)}, J^b(y)\right] = \delta ^{ab}\delta ^2_{S^2}(z-y).\label{jcom}
\eeq
In other words, our definition includes the suitable metrical factors which give the covariant $\delta$-function on the right hand side.

We now turn to the action of the kinetic energy operator on functionals involving two $J$ fields, such as $\int_{x,y}C^{m n}(x,y)J^m(x)J^n(y) $, where $C$ is some test function.
Mathematical manipulations, similar to what was carried out above,
show that
\beqar
\frac{e^2r^2}{2\pi }\int_{u,v,x,y}\!\!\!\!\!&&\!\!\!\!\Pi^{rs}(u,v)C^{mn}(x,y)[\bp^r(u),J^m(x][p^s(v),J^n(y)]\nonumber \\
&=&\frac{e^2Nr}{2\pi^2} \int_{u,v,x,y} C^{m n}(x,y)\left(\mD(v)\mg_-(u,v)\right)^{rs}\left[\frac{\delta}{\delta J^r(v)},J^m(x)\right] \left[\frac{\delta}{\delta J^s(u)},J^n(y)\right]
\nonumber\\
\label{kej10}
\eeqar
This shows that the term in $T$ involving two derivatives with respect to
$J^a$ is given by
 \beq T_2 = \frac{mr}{\pi}\int_{w,z}\left(\mD(w)\mg_-)^{ab}(z,w) \right)\frac{\delta }{\delta J^a(w)}
\frac{\delta }{\delta J^b(z)}\label{kej11}
\eeq

Putting together (\ref{t1}) and (\ref{kej11}), the expression for the kinetic energy operator in terms of the currents
is thus given by
\beq
T=  m\int_z J^a(z)\frac{\delta}{\delta J^a(z)} + \frac{mr}{\pi}\int_{w,z}\left(\mD(w)\mg_-)^{ab}(z,w) \right)\frac{\delta }{\delta J^a(w)}
\frac{\delta }{\delta J^b(z)}
\label{treg}
\eeq
with $\mD$ given by (\ref{kej2}).
If the two $J$ derivatives in $T$ act on well separated $J$ fields, then
$\mg_-$ can be replaced by  its unregulated version, so that
\beq
T=  m\int_z J^a(z)\frac{\delta}{\delta J^a(z)} + \frac{mr}{\pi}\int_{w,z}\mD^{ab}(w)G_-(z,w)\frac{\delta }{\delta J^a(w)}
\frac{\delta }{\delta J^b(z)}.\label{tunreg}
\eeq

With the use of the regularized $\delta$-function, the potential energy term can be written out in terms of currents as
\beqar
V &=&{\pi^2 \over mN r^2} \int_{z,w}  \left( \rmin J^a(z) [K(z,\bw ) K^{-1} (w, \bw)]^{ab} \rmin J^b\right)~ \sigma( z,w, \epsilon)\nonumber\\
&&\hskip .2in- {\pi \over Nr} \int_{z,w} \left[{\mathcal R}_{-w} {\mathcal R}_{-z} \mD_w^{ba}  G_{-}(z,w) \right]
[K(z,\bw ) K^{-1} (w, \bw)]^{ab} \sigma (z,w,\epsilon )
\label{kej12}
\eeqar
The second term on the right hand side is what needs to be subtracted to define a properly normal-ordered
expression.
\section{Vacuum Wave Functional}
In this section we utilize the Hamiltonian formalism developed above to compute the vacuum wave functional of the theory on ${\mathbb R}\times S^2$.
The analysis closely parallels the case of ${\mathbb R}\times {\mathbb R}^2$ discussed in \cite{KKNVWF}.
One of the motivations for the computation on  ${\mathbb R}\times S^2$ is to elucidate the  extent to which the physical results of the analysis on  ${\mathbb R}\times {\mathbb R}^2$ can be carried over to other spatial geometries.
In \cite{robustness}, it was argued that the value of the mass-gap and the functional form of the leading vacuum wave functional are
essentially determined by general features of the theory such as
Lorentz invariance and the two-dimensional anomaly. The explicit computation of the wave functional on ${\mathbb R}\times S^2$ will show that these arguments are indeed realized.

As in \cite{KKNVWF}, one takes an ansatz for the vacuum wave functional of the form
\beq \Psi_0 = e^P \label{vwf1}\eeq
where, $P$ is a functional of the $J$'s to be determined. The condition that this be a zero energy ground state of the theory translates  to the operator equation
\beq [T,P] + \frac{1}{2}[[T,P],P] + V =0\label{vwf-equation}\eeq
If the potential energy is neglected, since the kinetic energy involves derivatives with repsetc to $J$, a solution is evidently given by $P=0$ (or $\Psi_0 =1$, up to normalization).
The fact that $T$ is proportional to $m$ and $V$ to $1/m$ suggests that one can set up a ${1/m}$-expansion for P as
\beq P = \frac{c_0}{m^2} P_0 + \frac{c_1}{m^4}P_1 + \frac{c_2}{m^8}P_2 + \cdots\label{vwf2}
\eeq
Equation (\ref{vwf-equation}) then splits up as
\beqar
\frac{c_0}{m^2}[T,P_0] =V \hspace{2cm}\nonumber \\
\frac{c_1}{m^4}[T,P_1] + \frac{c_0}{2m^4}[T,[P_0,P_0]] =0, \cdots {\rm etc.}
\label{vwf3}
\eeqar
The leading contribution to the strong coupling wave functional is thus given by $P_0$.
In the planar case, a key relation
\beq [T, V ] = 2m V \label{tv}
\eeq
where $V$ was the properly normally-ordered expression, was the crucial ingredient for solving these equations  \cite{KKN1}. This relation
can be verified to be valid for the case of ${\mathbb R}\times S^2$ as well,
with the definition of the regularized $V$ as in (\ref{kej12}),
with the proper normal ordering term.
This implies that  $P_0 = -V/2m$, so that
\beq
\Psi_0 = \exp\left[- {\frac{\pi^2}{2m^2N}\int d\mu \left(\frac{\rmin}{r}J^a\frac{\rmin}{r}J^a\right) + \mathcal{O}(m^{-4})}\right].\label{vwf-j}
\eeq
Reverting back to the $A_{\pm}$ variables,
\beq \Psi_0 = \exp\left[-\frac{4}{2\pi m e^2}\int d\mu\mbox{Tr}\left(\frac{\sqrt{\pi }}{r}(\rpl A_- - \rmin A_+)+[A_+,A_-]\right)^2 + \mathcal{O}(m^{-4})\right].\label{vwf4}
\eeq
Evidently, these expressions reduce to the appropriate ones on the plane \cite{KKNVWF}, once the large $r$ limit is taken.

As expected, to this order, the vacuum wave functional is nothing but the action functional Yang-Mills theory defined on $S^2$. Thus, in complete analogy with the planar theory,  the vacuum expectation value of any spatial observable of the gauge theory can be re-cast as an appropriate (Euclidean) correlation function for the  two dimensional Yang-Mills theory in a functional integral framework.

The leading term in $P$ with two powers of the current and with arbitrary powers of momentum (or derivatives of $J$) can also be worked out as in the planar case.
This calculation is most easily phrased in terms of the momentum-space variables
$J^a_{l,m}$ which are the components of $J^a$ in a vector spherical harmonic expansion.
These expansions and the expressions for the kinetic and potential energy operators
are given in the next section.
Here we will use the scaled variable
\beq
I^a_{l,m} = \sqrt{\frac{l(l+1)}{2l+1}} J^a_{l,m}.
\eeq
To determine the leading term in the wave-functional,
we make the Gaussian ansatz for $P$,
\beq
P = P_G = \sum_{l,m} (-1)^m K(l) I^a_{l,m}I^a_{l,-m}\label{gauss-ansatz},
\eeq
where $K(l)$ is an as-yet-undetermined kernel.
Imposing (\ref{vwf-equation}) on the above ansatz and using the momentum-space representations of $T$ and $V$ (equations (\ref{momT}) and (\ref{momV}))
leads to the equation
\beq
\frac{4Nm}{\pi^2}K^2(l) -2mK(l) - \frac{\pi^2}{r^2mN} =0 \label{vwf5}
\eeq
As in the previous computation leading to (\ref{vwf4}) the subtraction of a normal ordering divergence is implied. The  solution  of  the above equation corresponding to a normalizable wave functional is given by
\beq
K(l) = -\frac{\pi^2}{Nmr^2}\left(\frac{1}{m + \sqrt{m^2 + \frac{4l(l+1)}{r^2}}}\right).
\eeq
Reverting back to the position space basis, we have
\footnote{In our convention, $R_-$ differs from the lowering operator on $SU(2)$ by a `-' sign.}
\beq
\Psi_0 \approx \Psi_G = \exp \left[-\frac{\pi^2}{Nm}\int d\mu \left(\frac{\rmin}{r}J^a\right)\frac{1}{m + \sqrt{m^2 - \frac{4\rmin R_+}{r^2}}}\left(\frac{\rmin}{r}J^a\right)\right]\label{vwf-rel}.
\eeq
It is interesting to check the large $r$ limit of this formula.
Using the correspondence of quantities on the sphere and the plane, outlined in
section 4,
we see that, as $r \rightarrow \infty $,
\beq
\Psi_G = \exp\left[ -\frac{\pi}{Nm}\int \bar{\partial} J^a \frac{1}{m + \sqrt{m^2 - 4\bar{\partial}\partial}}\bar{\partial} J^a\right]
\eeq
which agrees  completely with the results in \cite{KKNVWF} and the general arguments in \cite{robustness}.  It is also important to note that the presence  of the the additional parameter $r$ in turn generates a dimensionless parameter $rm$. The strong coupling limit of  (\ref{vwf-rel}) thus corresponds to taking $rm\gg 1$ while the reverse inequality gives us the weak coupling regime where perturbation theory is valid. It is straightforward to see that (\ref{vwf-rel}) interpolates smoothly between these two limits.

\section{Concluding Remarks}

In this paper we have extended the gauge-invariant Hamiltonian analysis of
(2+1)-dimensional Yang-Mills theories to the case where the spatial manifold is a two-sphere.

Our results could serve as theoretical predictions for lattice gauge theory computations of the mass-gap and the string tension of $SU(N)$ Yang-Mills theories on ${\mathbb R} \times S^2$. Other than lattice gauge theories, the thermodynamic properties of pure Yang-Mills theory on ${\mathbb R} \times S^3$ and ${\mathbb R} \times S^2$ have recently been investigated by a number of authors\cite{Aha-1, Aha-2}. The weakly coupled gauge theory, analyzed in \cite{Aha-2}, showed an interesting phase structure at high temperatures. Since most of the techniques that we developed in this paper naturally lend themselves to the analysis of the strong coupling regime of the gauge theory, it would be very interesting to extend the analysis to the finite temperature case and investigate the nature of the de-confining phase transition as a function of the radius of the sphere.

Perhaps the most interesting extension of the present analysis lies in the direction of supersymmetrization of the theory. In particular, analyzing the theory with sixteen supercharges  is of paramount importance for testing some very concrete string theory based predictions for the spectrum of the theory at strong coupling \cite{lin-maldacena}. We are in the process of analyzing this possibility. Several other fascinating results have also been conjectured for three dimensional Yang-Mills theories with diverse degrees of supersymmetry on ${\mathbb R}^3$\cite{seiberg-notes, seiberg-compactification, witten-index}. Analyzing these theories on ${\mathbb R}\times S^2$ using the methods presented in this paper also remains an interesting avenue for future explorations.

\vskip .1in
We are grateful to Dimitra Karabali, Prem Kumar and Alexios Polychronakos for various useful discussions.
This research was supported in part by the National Science Foundation
grant PHY-0555620 and by a PSC-CUNY grant.

\section*{\normalsize APPENDIX A: The Gauge-invariant Measure}
\def\theequation{A\arabic{equation}}
\setcounter{equation}{0}

As with Yang-Mills on $R^3$, the change of variables $A_+, A_- \rightarrow H$ involves a non-trivial Jacobian. The Jacobian is necessary
for computing the inner product on the space of wave functionals, which are taken to be functionals of $H$, or equivalently that of $J$. To compute the Jacobian we follow the analysis done on the plane quite closely. In what follows, we shall perform the relevant analysis on a sphere of unit volume i.e. at $r^2 =1$. The answer for a general value for $r^2$ can be obtained simply from dimensional analysis and it is mentioned at the very end of the section.

We first note that the distance functional on the space of the gauge potentials on $S^2$ can be written as\beq \delta s^2_A = -8\int d\mu(z) \mbox{Tr} (\delta A_+ \delta A_-) = 8\int d\mu(z) \mbox{Tr}\left((D_+\delta MM^{-1})(D_-M^{\dagger -1}\delta M^\dagger)\right),\eeq where $D_{\pm}$ are the covariant derivatives. Explicitly \beq D_+(\delta MM^{-1}) = \sqrt{\pi}R_+(\delta MM^{-1}) + [A_+,(\delta MM^{-1})].\eeq
The pre-factor $8$ is chosen so that the distance function goes over to that on $R^2$ once the sphere is de-compactified.

The measure on the space of $Sl(N,C)$ matrices $M$ is given by \beq \delta S^2_{Sl(N,C)} = 8\int d\mu(z) \mbox{Tr} (\delta MM^{-1})(M^{\dagger -1}\delta M^{\dagger}). \eeq Thus \beq d\mu(A) = \mbox{det}(D_+D_-)d\mu(M,M^{\dagger}).\eeq As expected, the jacobian is given by the Dirac determinant for massless Fermions on $S^2$. To evaluate the determinant, we first note that if we denote \beq S_+ = \ln\det D_+\eeq then its variation is given by \beq \delta S_+ = \int d\mu(x) \tr[D^{-1}_+(x,x)\delta A_+(x)].\eeq The variation requires the evaluation of the covariant Green's function at a coincident point, which of course requires a careful regularization. We shall proceed to evaluate this next. the unregulated version of $D^{-1}_+$ is given by \beq D^{-1}_+ (x,y) = \frac{M(x)(1+ \bar{x}y)M^{-1}(y)}{\sqrt{\pi}(\bar{x} - \bar{y})}.\eeq We regulate this expression as \beq D_+^{-1}(x,y) \rightarrow \mD^{-1}_+(x,y) = M(x)\left[\int_u\frac{(1-\bar{x}u)}{\sqrt{\pi}(\bar{x}-\bar{u})}K^{-1}(y,\bar{u})K(y,
\bar{y})\sigma(x,y,\epsilon)\right]M^{-1}(y).\eeq
The choice of regularization is by no means unique, and indeed one could have point split the unregulated expression in various different ways to construct its regularized version. A basic guiding principle to employ in the choice of regularizations is that the final answer be gauge-invariant, i.e expressible in terms of the variables $H$. We shall see that with our choice above, that will indeed be the case. Any other choice of regularization will lead to answers that will differ from our result by local counter-terms.

At coincident points, we can expand  $K^{-1}(x,\bar{u})$ about $\bar{u} =x$ to get \beqar \mD^{-1}_+(x,x) &=& \int_u \frac{1+\bar{x}u}{\sqrt{\pi}(\bar{x} - \bar{u})}\sigma(x,u,\epsilon) - \frac{1}{\sqrt{\pi}}[M(\partial_{\bar{x}}H)H^{-1}M^{-1}](x)\int_u (1+\bar{x}u)\sigma(x,u,\epsilon) +\mathcal{O}(\epsilon) \nonumber\\
& & = I_1 + I_2 + \mathcal{O}(\epsilon).\eeqar The first integral can be rewritten as \beq \sqrt{\pi}I_1 = -\int d\mu(u)G_+(u,x)\sigma(u,x,\epsilon) = -\int d\mu(g) G_+(g,g')\sigma(g,g',\epsilon)\eeq where $g,g'$ are the $SU(2)$ elements corresponding to $u$ and $x$ respectively. We now note that both $G_+ $ and $\sigma $ are functions of $g'^\dagger g$. Thus \beq -\sqrt{\pi}I_{1} =  \int d\mu(g) G_+(g,g')\sigma(g,g',\epsilon) = \int d\mu(g) G_+(g'^\dagger g)\sigma(g'^\dagger g,\epsilon) = \int d\mu(g) G_+(g')\sigma(g,\epsilon).\eeq In the last step we have used the left and right invariance of the integration measure on $SU(2)$. Thus, reverting back to the local coordinates on the sphere,
\beq -I_1 = \frac{1}{\sqrt{\pi}}\int d\mu(u)\frac{1}{\bar{u}}\sigma(u,\epsilon)\eeq which vanishes by angular integration. Thus \beq I_1 =0.\eeq  In $I_2$, which is devoid of short distance singularities, we can let $\epsilon $ approach zero inside the integrand, giving us the final result:\beq \mD^{-1}_+(x,x) = \frac{1}{\sqrt{\pi}}M(x)H^{-1}(x)(R_-H(x))M^{-1}(x) = (\frac{1}{\pi}A_- + \frac{1}{\sqrt{\pi}}(R_-M)M^{-1}).\eeq
Thus, putting together the analysis so far, \beq \delta S_+ = \int \tr((A_- + (R_-M)M^{-1})\delta A_+).\label{holvar}\eeq
To integrate this functional differential equation we note that $\delta S_+$ can be related to the holomorphic variation of a Hermitian WZW model on $S^2$. The  WZW action on $S^2$ is defined to be\beq S_{WZW}[A] = \frac{1}{2}\int d\mu(z) \tr(R_+AR_-A^{-1}) + \Gamma\label{WZW}\eeq where the volume term \beq \Gamma = \frac{i}{12\pi}\int d^3x \epsilon^{ijk}\tr(A^{-1}\partial_i A A^{-1}\partial_j A A^{-1}\partial_k A).\eeq The derivatives \beq \partial_1 = \frac{1}{1+z\bar{z}}(R_+ + R_-), \partial_2 = \frac{1}{i(1+z\bar{z})}(R_+ - R_-)\eeq while $x^3$ corresponds to the coordinate third direction whose boundary is $S^2$. We should note that despite their ostensible appearance all metrical factors eventually cancel out and we get back an action functional which is the same as that on $R^2$. This is to be expected from the toplogical nature of the WZW action.

The Polyakov-Wiegmann identity satisfied by (\ref{WZW}) can be written down as \beq S_{WZW}[AB] = S_{WZW}[A] + S_{WZW}[B] - \int d\mu(z) \tr(A^{-1}R_-AR_+BB^{-1}),\eeq which gives us the relation \beq S_{WZW}[H] = S_{WZW}[M] + S_{WZW}[M^\dagger ] + \frac{1}{\pi}\int d\mu(z) \tr(A_-A_+).\eeq
Thus the variation of the Wess-Zumino-Witten action with respect to the holomorphic component of the gauge potential is given by \beq \delta_{A_+} S_{WZW}[H] = (\frac{1}{\pi}A_- + \frac{1}{\sqrt{\pi}}(R_-M)M^{-1}).\label{holvar1}\eeq

The corresponding variation w.r.t the anti-holomorphic component of the gauge connection proceeds along exactly similar lines. Using (\ref{holvar},\ref{holvar1}) we have the functional differential equation \beq \delta_{A_+}S_+ = A_R \delta_{A_+}S_{WZW}[H].\label{holvar2}\eeq
$A_R$ is given by $\tr(t^at^b)_{R}= \tr(t^at^b)_{Fundamental}$. For us, $R$ corresponds to the adjoint representation of the group $G = SU(N)$.

Using the initial condition that, $D_{\pm} = \sqrt{\pi}R_{\pm}$ when $A_{\pm} =0$, we can integrate (\ref{holvar2}) and its anti-holomorphic counterpart, to get: \beq \det[D_+D_-] = \left[\frac{\det[\sqrt{\pi}R_+\sqrt{\pi}R_-]}{\int d\mu(z)}\right]^{\mbox{dim}(G)}\exp(2c_AS_{WZW}[H]).\label{measure}
\eeq
where $c_A$ is the quadratic Casimir invariant for the adjoint representation.
To obtain the result for a sphere of volume $r^2$, one simply replaces $R_{\pm} \rightarrow \frac{1}{r}R_{\pm}$ and the integration measure by $d\mu \rightarrow r^2 d\mu $.

\section*{\normalsize APPENDIX B: Mode Expansions}
\def\theequation{B\arabic{equation}}
\setcounter{equation}{0}

 In this appendix, we give the mode expansions for
 the current and for the kinetic and potential energy terms.

We start with the expansion of the current $J$ in spherical harmonics. From its very definition (\ref{sph27}), it is clear that it is a vector field on the sphere. It should hence be expanded in terms of vector spherical harmonics. An appropriate expansion is given by \beq J^a(g) = \frac{1}{r}\sum_{l=1, -l\leq m\leq l}^{l=\infty } J^a_{l,m} \mD^{l}_{m,1}(g).\label{jmod} \eeq This is indeed the correct expansion to consider as $J$ is obtained by $R_+$ acting on a scalar field on the sphere, hence its expansion involves $\mD^{l}_{m,1}$. Notice that
$\mD^{l}_{m,1}$ is not really a function on the sphere, as it involves the extra $U(1)$ direction as well. (This is as expected, since vectors are sections of the vector bundle.)
However, we shall see that in the final results the extra phases cancel out, even though they do show up at intermediate stages of the computation.

Using the orthogonality properties of the rotation matrices we can derive the following completeness relation
for the $\delta$-function appropriate to vectors.
\beq \delta (g',g) = \frac{1}{r^2}\sum_{l=1, -l\leq m\leq l}^{l=\infty } (2l+1) \mD^{*l}_{m,1}(g')\mD^l_{m,1}(g) = \frac{1}{r^2}\sum_{l=1, -l\leq m\leq l}^{l=\infty } (2l+1)\mD^l_{1,1}(g'^\dagger g).\eeq
This allows us to write\beq \frac{\delta}{\delta J^a(g)} =  \frac{1}{r}\sum_{l=1, -l\leq m\leq l}^{l=\infty } (2l+1) \mD^{*l}_{m,1}(g)\frac{\delta}{\delta J^a_{l,m}},\eeq which is consistent with (\ref{jcom}). Thus we can easily write
\beq T_1 = m\int d\mu(g) ~J^a(g)\frac{\delta}{\delta J^a(g)} = m\sum_{l=1, -l\leq m\leq l}^{l=\infty }J^a_{l,m}\frac{\delta}{\delta J^a_{l,m}}.\label{mom1}
\eeq
Notice that in the integral over $SU(2)$ in the left hand side of this
equation, the $U(1)$ factors cancel out and we eventually only perform an integral over the sphere. This will be true of the other manipulations we perform as well.
We now look at $T_2$, which we write in terms of $SU(2)$ coordinates as
\beq
T_2 =  \int_{g,g'} \left(m_1(R_+(g')G_-(g,g'))\frac{\delta^2}{\delta J^a(g)\delta J^a(g')}+ im_2f^{abc}G_-(g,g')J^c(g')\frac{\delta^2}{\delta J^b(g)\delta J^a(g')}\right),\label{t2su2}
\eeq
where,
\beq m_1 = \frac{mN}{\pi^2}, \hskip .2inm_2 = \frac{mr}{\pi}.
\eeq
Focussing on the first term in $T_2$ we  notice that \beq G_-(g,g') = R_+(g)G(g,g') \eeq where $G$ is the Green's function for the Laplacian with the mode expansion given by \beq r^2 G(g,g') = \sum_{l=1}^{\infty} \frac{2l+1}{l(l+1)}\mD^l_{0,0}(g'^\dagger g) = \sum_{l,m}\frac{2l+1}{l(l+1)}\mD^{*l}_{m,0}(g')\mD^l_{m,0}(g).\eeq
Thus, the first term in $T_2$ can be written as
\beqar
T^1_2 &=&
 \frac{m_1}{r^4} \int_{g,g'}\frac{(2l+1)(2p+1)(2q+1)}{l(l+1)}\nonumber\\
 &&\hskip .3in \times \left((R_+(g')R_+(g) \mD^{*l}_{m,0}(g')\mD^l_{m,0}(g))\mD^{*p}_{r,1}(g)\mD^{*q}_{s,1}(g')\right)\frac{\delta ^2}{\delta J^a_{p,r}\delta J^a_{q,s}}
 \label{t1modes}
\eeqar
Sum over the various momentum modes is implied in the above formula.
Summations will not be explicitly indicated to avoid cluttering the formulae.

The definition of the Wigner functions and $R_+$ give the relations
\beqar
R_+(g) \mD^l_{m,0}(g) &=& \sqrt{l(l+1)}\mD^l_{m,1}(g) \nonumber \\
 R_+(g') \mD^{*l}_{m,0}(g') &=& -\sqrt{l(l+1)}\mD^{*l}_{m,-1}(g')
 \eeqar
We can simplify (\ref{t1modes}) using these results as
  \beqar
  T^1_2 &=& -\frac{m_1}{r^4} \int_{g,g'}(2l+1)(2p+1)(2q+1)\left( \mD^{*l}_{m,-1}(g')\mD^l_{m,1}(g)\mD^{*p}_{r,1}(g)\mD^{*q}_{s,1}(g')\right)\frac{\delta ^2}{\delta J^a_{p,r}\delta J^a_{q,s}}\nonumber \\
& & = -\frac{m_1}{r^2}  \int_{g'}(2p+1)(2q+1)\left[\mD^{*l}_{m,-1}(g')\mD^{*q}_{s,1}(g')\right]\frac{\delta ^2}{\delta J^a_{l,m}\delta J^a_{q,s}} \nonumber \\
& & = m_1(-1)^{s+1}(2l+1)\frac{\delta ^2}{\delta J^a_{l,-s}\delta J^a_{l,s}}.
 \eeqar

Turning now to the second term in $T_2$, we can write it out as
\beqar
T_2^2 &=& im_2f^{abc}\int_{z,w}J^c(w) G_(z,w)\frac{\delta}{\delta J^b(z)}\frac{\delta}{\delta J^a(w)}\hspace{4cm}\nonumber \\
&=&i\frac{m}{\pi}f^{mnc}\int_{g,g'}\frac{(2r+1)(2a+1)(2l+1)}{l(l+1)}\nonumber\\
&&\hskip .5in \times \left( R_+(g)\mD ^{*l}_{q,0}(g')\mD ^{l}_{q,0}(g)\right)\mD^{p}_{q,1}(g')\mD ^{*r}_{s,1}(g)\mD ^{*a}_{b,1}(g')\left[J^{c}_{p,q}\frac{\delta}{\delta J^n_{r,s}}\frac{\delta}{\delta J^{m}_{a,b}}\right]\nonumber \\
&=& i\frac{m}{\pi}f^{mnc}\int_{g'}\frac{(2r+1)(2a+1)}{\sqrt{r(r+1)}}\mD^{p}_{q,1}(g')\mD ^{*r}_{s,0}(g')\mD ^{*a}_{b,1}(g')\left[J^{c}_{p,q}\frac{\delta}{\delta J^n_{r,s}}\frac{\delta}{\delta J^{m}_{a,b}}\right]
\eeqar
The last two integrals above are meant to be evaluated on a sphere (or equivalently an $SU(2)$) of unit volume. All the factors of $r$ cancel out.
The integral of three Wigner functions can be readily expressed in terms of Clebsch-Gordan coefficients. To do that we first note the that \beq \mD^l_{m,0}(\theta, \phi) = \sqrt{\frac{4\pi}{2l+1}}Y^*_{l,m}(\theta,\phi) \eeq where the usual spherical harmonics are normalized to unity on a sphere of volume $4\pi$. i.e. \beq \int \sin(\theta)d\theta d\phi Y^*_{l,m}(\theta, \phi)Y_{l',m'}(\theta,\phi) = \delta_{l,l'}\delta_{m,m'}.\eeq
It then follows that \beq \mD^l_{m,n}(\theta, \phi, \psi) = e^{-in\psi}\mD^l_{m,0}(\theta, \phi) =e^{-in\psi} \sqrt{\frac{4\pi}{2l+1}}Y^*_{l,m}(\theta,\phi).\eeq
Hence \beq \int_{g'}\mD^{p}_{q,1}(g')\mD ^{*r}_{s,0}(g')\mD ^{*a}_{b,1}(g') = \sqrt{\frac{4\pi}{(2p+1)(2a+1)(2r+1)}}\int d\tilde \Omega Y^*_{p,q}(\tilde \Omega) Y_{a,b}(\tilde \Omega) Y_{r,s}(\tilde \Omega),\eeq where $d\tilde \Omega = \sin(\theta)d\theta d\phi$, the volume measure on a sphere of volume $4\pi$. This last integral is a standard one, and can be evaluated to get the final answer as \beq \int_{g'}\mD^{p}_{q,1}(g')\mD ^{*r}_{s,0}(g')\mD ^{*a}_{b,1}(g') = \frac{1}{2p+1}C(r,a,p;0,0,0)C(r,a,p;s,b,q).\eeq In terms of the Wigner $3j$-symbol, \beq \int_{g'}\mD^{p}_{q,1}(g')\mD ^{*r}_{s,0}(g')\mD ^{*a}_{b,1}(g') = \left(
                                                                                                            \begin{array}{ccc}
                                                                                                              r & a & p \\
                                                                                                              0 & 0 & 0 \\
                                                                                                            \end{array}\right)\left(
                                                                                                                         \begin{array}{ccc}
                                                                                                                           r & a & p \\
                                                                                                                           s & b & -q \\
                                                                                                                         \end{array}
                                                                                                                       \right).
\eeq
Thus putting it all together, \beq T_2^2 = i\frac{m}{\pi}f^{mnc}\frac{(2r+1)(2a+1)}{\sqrt{r(r+1)}}
\left(\begin{array}{ccc}
r & a & p \\
0 & 0 & 0 \\
\end{array}\right)\left(
\begin{array}{ccc}
r & a & p \\
s & b & -q \\
\end{array}
\right)\left[J^{c}_{p,q}\frac{\delta}{\delta J^n_{r,s}}\frac{\delta}{\delta J^{m}_{a,b}}\right]\eeq
Hence, we finally have:\beqar T = m \left( J^a_{l,s}\frac{\delta}{\delta J^a_{l,s}} + \frac{N}{\pi^2}(-1)^{s+1}(2l+1) \frac{\delta ^2}{\delta J^a_{l,s} \delta J^a_{l,-s}}\right)\hspace{2.7cm}\nonumber \\
+\frac{im}{\pi}f^{mnc}\frac{(2r+1)(2a+1)}{\sqrt{r(r+1)}}
\left(\begin{array}{ccc}
r & a & p \\
0 & 0 & 0 \\
\end{array}\right)\left(
\begin{array}{ccc}
r & a & p \\
s & b & -q \\
\end{array}
\right)J^{c}_{p,q}\frac{\delta}{\delta J^n_{r,s}}\frac{\delta}{\delta J^{m}_{a,b}}.\label{momT}
\eeqar
Once again, sum over the repeated momentum indices is implied.

The mode expansion of the potential energy term given in (\ref{sph26}) is easily worked out as
\beq
V = {2\pi^3 \over e^2 N^2 r^2} \sum_{l\geq 1}   {l(l+1) \over 2l+1} \sum_m (-1)^m
J^a_{l,m} J^a_{l,-m}
\label{momV}
\eeq


\begin{thebibliography}{99}
\bibitem{KKN1}
  D.~Karabali and V.~P.~Nair,
  ``A gauge-invariant Hamiltonian analysis for non-Abelian gauge theories in
  (2+1) dimensions,''
  Nucl.\ Phys.\  B {\bf 464}, 135 (1996)
  [arXiv:hep-th/9510157];

   D.~Karabali and V.~P.~Nair,
  ``On the origin of the mass gap for non-Abelian gauge theories in (2+1)
  dimensions,''
  Phys.\ Lett.\  B {\bf 379}, 141 (1996)
  [arXiv:hep-th/9602155];

  D.~Karabali, C.~j.~Kim and V.~P.~Nair,
  ``Planar Yang-Mills theory: Hamiltonian, regulators and mass gap,''
  Nucl.\ Phys.\  B {\bf 524}, 661 (1998)
  [arXiv:hep-th/9705087];

 D.~Karabali, C.~j.~Kim and V.~P.~Nair,
  ``gauge-invariant variables and the Yang-Mills-Chern-Simons theory,''
  Nucl.\ Phys.\  B {\bf 566}, 331 (2000)
  [arXiv:hep-th/9907078].

\bibitem{KKNVWF}
  D.~Karabali, C.~j.~Kim and V.~P.~Nair,
  ``On the vacuum wave function and string tension of Yang-Mills theories  in
  (2+1) dimensions,''
  Phys.\ Lett.\  B {\bf 434}, 103 (1998)
  [arXiv:hep-th/9804132].

  \bibitem{glueballs1}
  R.~G.~Leigh, D.~Minic and A.~Yelnikov,
  ``On the Glueball Spectrum of Pure Yang-Mills Theory in 2+1 Dimensions,''
  Phys.\ Rev.\  D {\bf 76}, 065018 (2007)
  [arXiv:hep-th/0604060].

\bibitem{AKN}
  A.~Agarwal, D.~Karabali and V.~P.~Nair,
  ``Yang-Mills Theory in 2+1 Dimensions: Coupling of Matter Fields and
  String-breaking Effects,''
  Nucl.\ Phys.\  B {\bf 790}, 216 (2008)
  [arXiv:0705.0394 [hep-th]].

 \bibitem{Abe}
  Y.~Abe,
  ``On the deconfining limit in (2+1)-dimensional Yang-Mills theory,''
  arXiv:0804.3125 [hep-th].




  \bibitem{teper-1}
  M.~J.~Teper,
  ``SU(N) gauge theories in 2+1 dimensions,''
  Phys.\ Rev.\  D {\bf 59}, 014512 (1999)
  [arXiv:hep-lat/9804008];

    B.~Lucini and M.~Teper,
  ``SU(N) gauge theories in 2+1 dimensions: Further results,''
  Phys.\ Rev.\  D {\bf 66}, 097502 (2002)
  [arXiv:hep-lat/0206027];

  B.~Bringoltz and M.~Teper,
  ``A precise calculation of the fundamental string tension in SU(N) gauge
  theories in 2+1 dimensions,''
  Phys.\ Lett.\  B {\bf 645}, 383 (2007)
  [arXiv:hep-th/0611286].

  \bibitem{robustness}
  D.~Karabali and V.~P.~Nair,
  ``The robustness of the vacuum wave function and other matters for
  Yang-Mills theory,''
  Phys.\ Rev.\  D {\bf 77}, 025014 (2008)
  [arXiv:0705.2898 [hep-th]].

\bibitem{QH-1}
  D.~Karabali and V.~P.~Nair,
  ``Quantum Hall effect in higher dimensions,''
  Nucl.\ Phys.\  B {\bf 641}, 533 (2002)
  [arXiv:hep-th/0203264];

  D.~Karabali and V.~P.~Nair,
  ``The effective action for edge states in higher dimensional quantum Hall
  systems,''
  Nucl.\ Phys.\  B {\bf 679}, 427 (2004)
  [arXiv:hep-th/0307281];

  D.~Karabali, V.~P.~Nair and S.~Randjbar-Daemi,
  ``Fuzzy spaces, the M(atrix) model and the quantum Hall effect,''
  arXiv:hep-th/0407007.



\bibitem{Aha-1}
  O.~Aharony, J.~Marsano, S.~Minwalla, K.~Papadodimas and M.~Van Raamsdonk,
  ``A first order deconfinement transition in large N Yang-Mills theory on  a
  Phys.\ Rev.\  D {\bf 71}, 125018 (2005)
  [arXiv:hep-th/0502149].

\bibitem{Aha-2}
  K.~Papadodimas, H.~H.~Shieh and M.~Van Raamsdonk,
  ``A second order deconfinement transition for large N 2+1 dimensional
  Yang-Mills theory on a small S**2,''
  JHEP {\bf 0704}, 069 (2007)
  [arXiv:hep-th/0612066].


\bibitem{lin-maldacena}
  H.~Lin and J.~M.~Maldacena,
  ``Fivebranes from gauge theory,''
  Phys.\ Rev.\  D {\bf 74}, 084014 (2006)
  [arXiv:hep-th/0509235];

  J.~M.~Maldacena, M.~M.~Sheikh-Jabbari and M.~Van Raamsdonk,
  ``Transverse fivebranes in matrix theory,''
  JHEP {\bf 0301}, 038 (2003)
  [arXiv:hep-th/0211139].

\bibitem{seiberg-notes}
  N.~Seiberg,
  ``Notes on theories with 16 supercharges,''
  Nucl.\ Phys.\ Proc.\ Suppl.\  {\bf 67}, 158 (1998)
  [arXiv:hep-th/9705117].

\bibitem{seiberg-compactification}
  N.~Seiberg and E.~Witten,
  ``Gauge dynamics and compactification to three dimensions,''
  arXiv:hep-th/9607163.

\bibitem{witten-index}
  E.~Witten,
  ``Supersymmetric index of three-dimensional gauge theory,''
  arXiv:hep-th/9903005.


\end{thebibliography}
\end{document}